\documentclass[aps,floats,twocolumn,showpacs]{revtex4-1}
\usepackage[english]{babel}
\usepackage{amssymb}
\usepackage{amsmath}
\usepackage{graphicx}
\usepackage{cmap}

\usepackage{hyperref}
\hypersetup{colorlinks=true,urlcolor=blue,citecolor=blue,linkcolor=blue}

 \def\bc{\begin{center}}          \def\ec{\end{center}}
 \allowdisplaybreaks
 
\begin{document}
 \title{Exact kinetic theory for the instability of an electron beam in a hot magnetized plasma}
 \author{I.V. Timofeev and V.V. Annenkov}
 \affiliation{Budker Institute of Nuclear Physics SB RAS, Novosibirsk, Russia \\
 Novosibirsk State University, Novosibirsk, Russia}
 \date{\today}
 \begin{abstract}
Efficiency of collective beam-plasma interaction strongly depends on the growth rates of dominant instabilities excited in the system. Nevertheless, exact calculations of the full unstable spectrum in the framework of relativistic kinetic theory for arbitrary magnetic fields and particle distributions were unknown until now. In this paper we give an example of such a calculation answering the question whether the finite thermal spreads of plasma electrons are able to suppress the fastest growing modes in the beam-plasma system. It is shown that nonrelativistic temperatures of Maxwellian plasmas can stabilize only the oblique instabilities of relativistic beam. On the contrary, non-Maxwellian tails typically found in laboratory beam-plasma experiments are able to substantially reduce the growth rate of the dominant longitudinal modes affecting the efficiency of turbulent plasma heating.
 \end{abstract}
 \maketitle

\section{Introduction}
Collective beam-plasma interaction is one of the most long-standing and fundamental problems in plasma physics. Excitation of plasma oscillations by electron beams plays an important role in some astrophysical phenomena such as gamma-ray bursts \cite{Piran2005}, generation of high-energy cosmic rays \cite{Blandford1987}, type III solar radio bursts \cite{Gurnett1976,Li2006}, as well as in laboratory experiments designed to achieve the fusion ignition conditions in open magnetic systems \cite{Burdakov2010} and in inertial confinement fusion \cite{Tabak1994}.    

One of the most important characteristics of beam-plasma interaction is the linear growth rate of oscillations driven in an unstable beam-plasma system. This quantity not only determines  the rate of exponential growth of the wave energy at the linear stage of instability, but strongly affects the level of its nonlinear saturation.  For instance, if the growing mode with the wavenumber $k$ is stabilized by beam trapping, its saturation energy $W$ can be found from the equality of the instability growth rate $\Gamma$ and the bounce frequency $\omega_b$ of trapping electrons:
\begin{equation}
\Gamma \propto \omega_b=\sqrt{\frac{e k E}{m_e}}, \quad W\propto E^2 \propto \Gamma^4,	
\end{equation}
where $e$ and $m_e$ are the charge and mass of electron, and $E$ is the amplitude of electric field in such wave.
Simulations of beam-plasma interaction for the conditions typical to open traps \cite{Timofeev2010} show that long-time evolution of the beam-driven plasma turbulence comes finally to the regime of the constant pump power \cite{Timofeev2006}  that is saturated at the level corresponding to the emergence of beam trapping.
It means that the power pumping by an electron beam into a plasma turbulence in this regime is very sensitive to the typical growth rate of the most unstable resonant modes, $P\propto \Gamma W \propto \Gamma^5$, and that is the reason for accurate kinetic calculation of $\Gamma$.

The linear analysis of possible instabilities in the beam-plasma system is a subject of interest in a number of papers (see the review \cite{Bret2010a}), the first of which were published in 1940s \cite{Akhiezer1949,Bohm1949}. Unfortunately, most of them are based on various simplified models in which the linear response of beam and plasma electrons is calculated by using either the fluid or nonrelativistic Vlasov equations. It is also common to simplify the theory by considering specific directions of wave propagation \cite{Tautz2005a,Tautz2006} or by using some {\it a priori} ideas about polarizations of unstable waves. The first calculations of the full unstable spectrum taking into account both kinetic and relativistic effects without the use of any simplifying assumptions were performed just recently for the case of unmagnetized plasma with the Maxwell-J$\ddot{\mbox{u}}$ttner distribution function \cite{Bret2008,Bret2010}. 

In magnetized plasmas it is exceedingly difficult to obtain solutions of dispersion equation for the beam-plasma system, and that is why magnetic field effects have been initially studied  for diluted cold beams and cold plasmas \cite{Godfrey1975} and later for hot electron streams with arbitrary density ratios within the fluid approach \cite{Bret2006}. Calculation of the full unstable spectrum in the magnetized beam-plasma system  in the framework of relativistic kinetic theory requires large amount of computing resources and is referred to in the literature as the Clemmow-Dougherty's "daunting task" \cite{Clemmow1990,Bret2010a}. The main challenge is due to the fact that each component of the dielectric tensor for a hot magnetized plasma with an axially symmetric distribution function contains slowly convergent infinite series of products of Bessel functions, which must be also integrated twice over momenta. The first step toward the complete solution of this problem have been made in Ref. \cite{Timofeev2009} where the combined influence of kinetic and magnetic effects on the instability growth rate was studied for the case of angularly spread monoenergetic electron beam in a cold plasma.

The goal of this paper is to obtain the complete numerical solution of the Clemmow-Dougherty's task for arbitrary distribution functions of beam and plasma electrons. From the pure physical point of view we are interested in the question of how effectively the thermal spreads of plasma electrons can suppress the instability of the fastest growing modes. Our interest to this question is motivated by the laboratory experiments at the GOL-3 facility \cite{Burdakov2010} where turbulent plasma heating is accompanied by formation of strongly non-Maxwellian momentum distributions with the typical tails of superthermal electrons ($f\propto p^{-5}$) containing most of the plasma kinetic energy. It is obvious that formation of such an intense tail during beam injection can significantly reduce the growth rate of instability or even completely stabilize the system, making the strategy of increasing the plasma temperature by increasing the beam duration ineffective. Since in the modern concept of a fusion reactor \cite{Burdakov2011} based on open magnetic systems the long-pulse electron beams play a key role, the study of kinetic effects on the excitation efficiency of plasma turbulence is of particular interest.

\section{Formulation of the problem for the numerical study}

The main difficulties in numerical analysis of unstable oscillations in the magnetized beam-plasma system arise from  the infinite series
\begin{equation}\label{rd}
    S(a,z)=\sum\limits_{n=-\infty}^{\infty}
    \frac{J_n^2(z)}{a-n}
\end{equation}
that converges very slowly for the large arguments of Bessel functions. It was found that such a series can be summed \cite{Lerche1966,Newberger1982} and expressed in terms of Bessel functions with real arguments $z$ and complex orders $a$:
 \begin{equation}
    S(a,z)=\frac{\pi}{\sin \pi a} J_{-a}(z)J_a(z).
\end{equation}
This summation rule was used in Ref. \cite{Qin2007,Schlickeiser2010,Tautz2012} to reduce the dielectric tensor to some new alternative forms, which can simplify the  numerical solution of the magnetized plasma problem. Thus, it is convenient to write down the dielectric tensor for the magnetized beam-plasma system with hot electron components in the simplified form:
\begin{equation}\label{diel}
    \varepsilon_{\alpha\beta}=\delta_{\alpha\beta}+\chi^{(e)}_{\alpha\beta}+ \chi^{(b)}_{\alpha\beta},
\end{equation}
\begin{multline}
    \chi^{(\sigma)}_{\alpha\beta}=\frac{2
    \pi}{\omega^2}\int\limits_{-\infty}^{\infty} dp_{\parallel}
    \int\limits_{0}^{\infty} dp_{\perp} p_{\perp} \times \\
		\left[\frac{\omega v_{\parallel}}{\omega-k_{\parallel}
    v_{\parallel}} \frac{\partial f^{(\sigma)}}{\partial p_{\parallel}} h_{\alpha} h_{\beta}+ V
    T_{\alpha\beta}\right],
\end{multline}
where
 \begin{align}
&V=\frac{p_{\perp}}{p_{\parallel}}\left[v_{\parallel} \frac{\partial f^{(\sigma)}}{\partial p_{\perp}} -v_{\perp}
\frac{\partial f^{(\sigma)}}{\partial p_{\parallel}}\left(1-
\frac{\omega}{\omega -k_{\parallel}
 v_{\parallel}}\right)\right], \\
    &T_{xx}= \frac{a^2}{z^2} \left(R G_a-1\right), \\
    &T_{yy}= \frac{a^2}{z^2} -\frac{R}{2} \left(G_{a+1}+G_{a-1} +2 \frac{a^2}{z^2} G_a
    \right), \\
    &T_{xy}= -T_{yx}=-i  \frac{R}{4}
    \left(G_{a+1}-G_{a-1}\right), \\
    &T_{xz}= T_{zx}=\frac{p_{\parallel}}{p_{\perp}}\frac{a}{z} \left(R G_a-1\right), \\
    &T_{yz}= -T_{zy}=i \frac{R}{4} \frac{p_{\parallel}}{p_{\perp}}\frac{z}{a} \left(G_{a+1}-G_{a-1}\right),
    \\
    &T_{zz}= \frac{p_{\parallel}^2}{ p_{\perp}^2}
    \left(R G_{a}-1\right), \\
		&R= \frac{\pi a}{\sin \pi a}, \qquad z=\frac{k_{\perp} p_{\perp}}{\Omega}, \qquad
a=\frac{\gamma\omega -k_{\parallel} p_{\parallel}}{\Omega}, \nonumber \\
&G_a= J_{-a}(z) J_a(z),\qquad h_{\alpha}=\frac{B_{\alpha}}{B}, \qquad \Omega=\frac{e B}{m_e c \omega_p}. \nonumber
\end{align}
Here the uniform magnetic field ${\bf B}=(0, 0, B)$ is aligned with the beam velocity, the wave frequency $\omega$ is expressed in units of plasma frequency $\omega_p$, the wavevector ${\bf k}=(k_{\bot}, 0, k_{||})$ in units of $\omega_p/c$, velocities ${\bf v}$ are measured in the speed of light $c$, and momenta ${\bf p}=\gamma {\bf v}$ in units of $m_e c$.

Distribution functions for the beam and plasma electrons are normalized to their own relative densities
\begin{equation}\label{norm}
    \int f^{(\sigma)} d^3 p=n^{(\sigma)}
\end{equation}
and satisfy the conditions 
\begin{equation}
    \sum\limits_\sigma \int {\bf v} f^{(\sigma)} d^3 p=0, \quad n^{(p)}+n^{(b)}=1.
\end{equation}
\begin{figure*}[htb]
\bc\includegraphics[width=440bp]{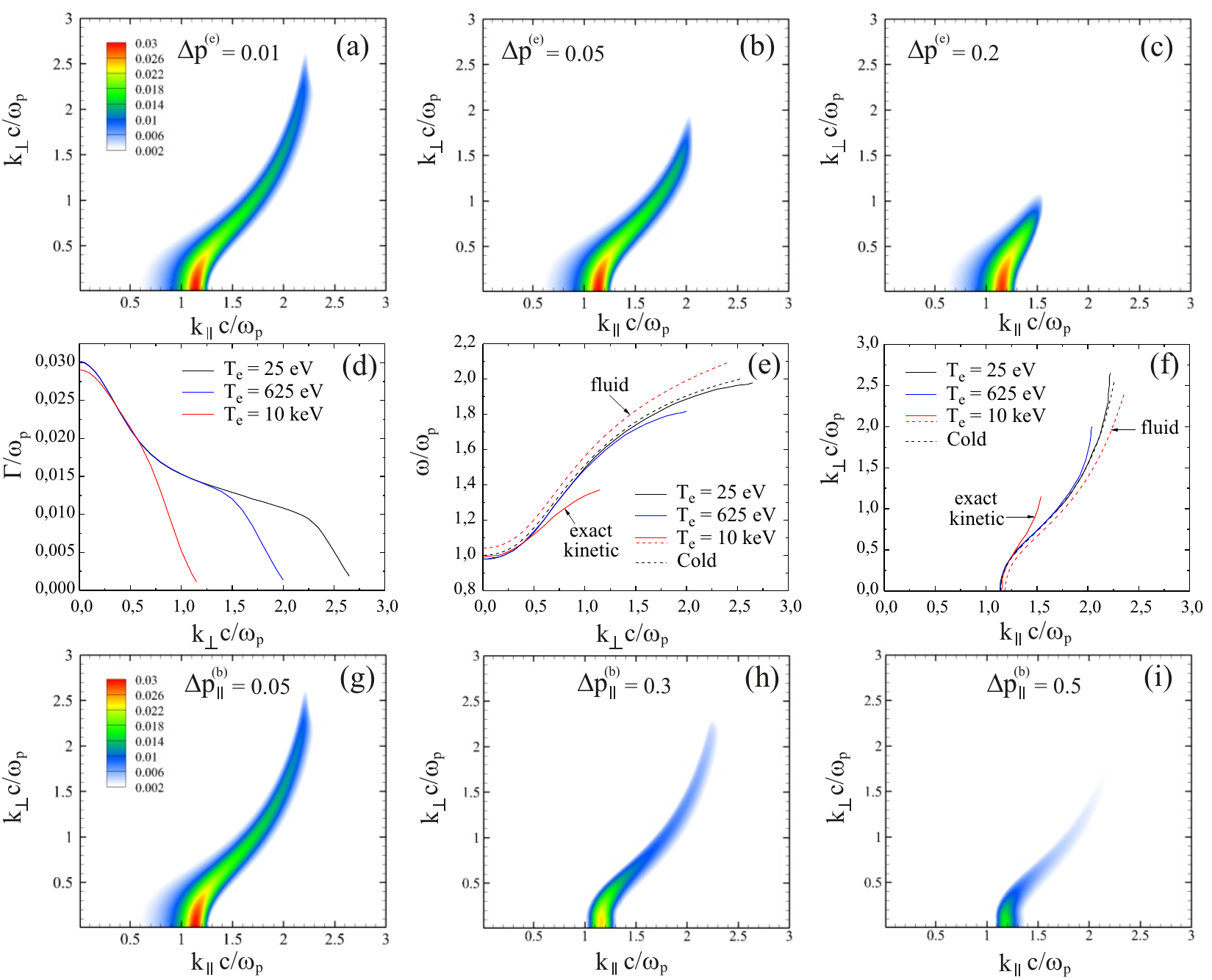} \ec \caption{Kinetic effects in Maxwellian plasmas. (a), (b) and (c) Transformation of the growth rate map $\Gamma(k_{\bot},k_{||})$ with the increase of plasma electron temperature for the fixed beam parameters $\Delta p_{||}^{(b)}=0.05$ and $\Delta
p_{\bot}^{(b)}=0.5$. (d) and (e) The imaginary and real parts of the frequency of unstable waves along the lines $k_{||}=k_{||}(k_{\bot})$ of maximal growth rate for the maps (a), (b) and (c). (f) The form and position of $k$-lines.  (g), (h) and (i). Supression of the beam-plasma instability with the rise of longitudinal momentum beam spread for the fixed $\Delta p^{(e)}=0.01$ and $\Delta
p_{\bot}^{(b)}=0.5$. The magnetic field $\Omega=2$.}\label{om2}
\end{figure*}
Effects of mobile ions do not make the problem more complicated, but their influence on the stabilty of high-frequency oscillations is negligibly small and is not taken into account.

Despite the fact that the infinite series $S(a,z)$ could be expressed in terms of Bessel functions $J_a(z)$, computation of this value remains to be a difficult task. The reason for that is when the growth rate is calculated for relativistic distributions in a wide wavenumber space, the order and argument of Bessel function can take large values simultaneously. It means that neither power expansions of  $J_a(z)$, nor the well known asymptotic formulas for large $z$ cannot be used for efficient computations in this case. Analysis shows that the most universal algorithm that works efficiently throughout the whole $(a,z)$-space is based on the following integral representation:
 \begin{equation}
    G_a=\frac{e^{i \pi a}}{2 \pi} \int\limits_{0}^{2 \pi} e^{-i a
    \varphi} J_0\left(2z\sin\frac{\varphi}{2}\right) d\varphi.
\end{equation}
The same representation does also arise when the dielectric tensor is derived by the novel method \cite{Qin2007} based on the symmetry of electron trajectories in the magnetic field.

Effects of finite thermal spreads of plasma and beam electrons are studied here for the anisotropic Maxwellian
\begin{multline}\label{max}
    f^{(\sigma)}({\bf p})= C_1 \exp\left[-\frac{p_{\perp}^2}{\Delta p_{\perp}^{(\sigma)2}}
    -\frac{(p_{\parallel}-p^{(\sigma)})^2}{\Delta
    p_{\parallel}^{(\sigma)2}}\right] \times \\
		\left[H(p_{||}-p^{(\sigma)}+3 \Delta p_{||}^{(\sigma)})-H(p_{||}-p^{(\sigma)}-3 \Delta p_{||}^{(\sigma)})\right] \times \\
		H(3\Delta p_{\bot}^{(\sigma)}-p_{\bot}),
\end{multline}
and for the strongly non-Maxwellian distribution containing the energetic power tail that is typically formed in the beam-plasma experiments \cite{Astrelin1997},
  \begin{multline}\label{e1}
    f^{(e)}({\bf p})= \frac{C_2 H(p_h-p_{\bot}) }{(p_{\bot}^2 +(p_{||}-p^{(e)})^2+\Delta p^{(e) 2})^{5/2}}\times \\
		\left[H(p_{||}-p^{(e)}+p_h)-H(p_{||}-p^{(e)}-p_h)\right],
\end{multline}
where $H(p)$ is the step function, $p_h$ is the threshold value of momentum components above which there are no particles in the plasma distribution, $\Delta p ^{(\sigma)}$ are the typical momentum spreads of beam and plasma electrons, $p ^{(\sigma)}$ are the mean drift momenta, and $C_1$, $C_2$ are the coefficients corresponding to normalizations (\ref{norm}). 
In the isotropic Maxwellian plasma the temperature of plasma electrons for small momentum spreads can be  approximately calculated as $T_e/m_e c^2=\Delta p^{(e) 2}/2$. In the case of strongly non-Maxwellian distributions, this quantity loses its conventional meaning \cite{Timofeev2013}, since the kinetic energy concentrated in superthermal electrons significantly exceeds the energy of the distribution "core".

\section{Computation results}
\subsection{Maxwellian plasmas}
First, let us study the influence of finite electron temperature in the Maxwellian plasma on the instability driven by the electron beam with the anisotropic distribution (\ref{max}) ($\Delta p_{||}^{(b)}=0.05$ and $\Delta p_{\bot}^{(b)}=0.5$),  relative density $n^{(b)}=0.002$, and averaged momentum $p^{(b)}=2.06$ in the magnetic field  $\Omega=2$. 
These parameters are very close to those chosen in Ref. \cite{Timofeev2009} for the monoenergetic beam, but in our case the transition to the smoother momentum distribution results in stabilization of various weak instabilities driven on the cyclotron resonances. That is the reason why we observe only the Cherenkov buildup of the upper-hybrid mode on the further presented growth rate maps.

As one can see in Fig. \ref{om2} (a), (b) and (c), the increase in plasma temperature suppresses oblique instabilities and has almost no impact on the unstable waves propagating along the magnetic field. It means that heating of plasma electrons up to the temperatures $\sim 10$ keV during the beam injection  cannot stabilize the most unstable modes in the system. These modes, however, appear to be much more sensitive to the beam thermal spreads. Fig. \ref{om2} (g), (h) and (i) show that in contrast to the previous case the instability growth rate decreases uniformly in the whole $k$-space  as the longitudinal momentum spread of the beam electrons increases.
\begin{figure*}[htb]
\bc\includegraphics[width=450bp]{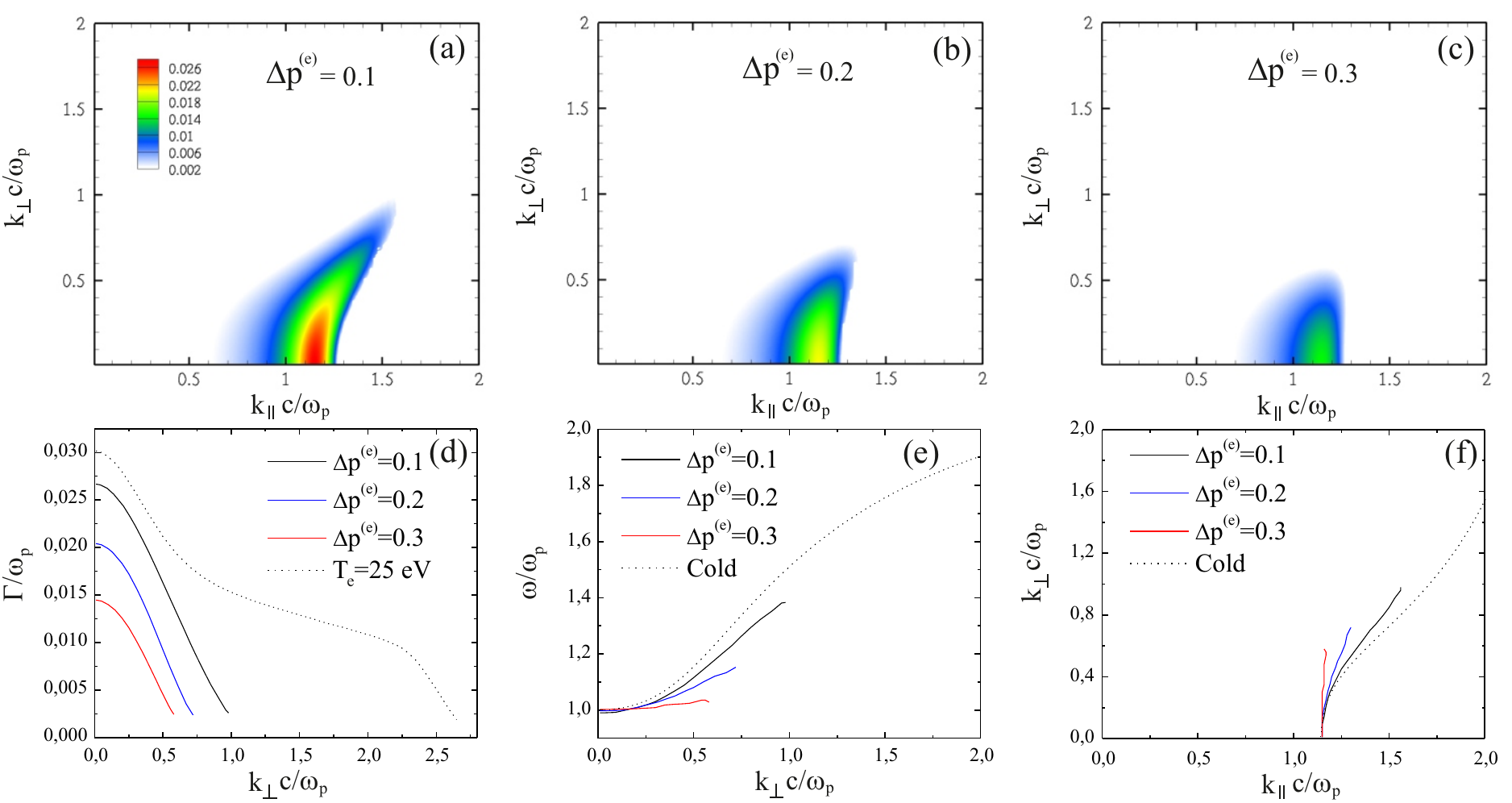} \ec \caption{Kinetic effects in nonmaxwellian plasmas. (a), (b), (c) Transformation of the growth rate map $\Gamma(k_{\bot},k_{||})$ with the increase of thermal spreads of plasma electrons for the fixed beam parameters $\Delta p_{||}^{(b)}=0.05$ and $\Delta p_{\bot}^{(b)}=0.5$. (d), (e) The imaginary and real parts of frequency of unstable waves along the lines $k_{||}=k_{||}(k_{\bot})$ of maximal growth rate for the case (a), (b), (c). (f) The form and position of $k$-lines.  The magnetic field and threshold momentum $\Omega=2$ and $p_h=5$.}\label{r2}
\end{figure*}

The stabilizing role of plasma temperature is visible more clearly in Fig. \ref{om2} (d) and (e), where the real and imaginary parts of wave frequency are calculated along the lines $k_{\bot} (k_{||})$ showing the position of local maximum of the growth rate in the $k$-space. For weak beams, the form of such a line can be found from coupling of the dispersion curves describing the proper plasma waves $\omega_k (k_{\bot},k_{||})$ and the beam branches at frequencies $k_{||} v_b +n\Omega/\gamma$.
It is seen that the finite temperature of plasma electrons results in appearance of the spectral region with a rather sharp decline of the growth rate, which moves to small $k_{\perp}$ with the increase of the thermal spread. One can also see the decrease in the real frequency of unstable waves inside this region in comparison to the case of cold plasma. It contradicts the fluid theory predicting the positive frequency shift for the upper-hybrid mode in the hot magnetized plasma. Indeed, in the fluid approach accounting for the finite electron pressure, the dispersion of this mode is governed by the dielectric tensor:
\begin{align}
\varepsilon_{xx} &= 1-A \left(1-\frac{k_{\parallel}^2
V_T^2}{\omega^2}\right), \\
\varepsilon_{xy} &= -\varepsilon_{xy}=i \frac{\Omega}{\omega} A
\left(1-\frac{k_{\parallel}^2 V_T^2}{\omega^2}\right), \\
\varepsilon_{yy} &= 1-A \left(1-\frac{k^2 V_T^2}{\omega^2}\right),\\
\varepsilon_{xz} &= \varepsilon_{zx}= -A \frac{k_{\parallel}
k_{\perp} V_T^2}{\omega^2}, \\
\varepsilon_{yz} &= -\varepsilon_{zy}=-i \frac{\Omega}{\omega} A
\frac{k_{\parallel} k_{\perp} V_T^2}{\omega^2}, \\
\varepsilon_{zz} &= 1-A \left(1-\frac{k_{\perp}^2
V_T^2+\Omega^2}{\omega^2}\right), \\
A &= \left(\omega^2-\Omega^2-k^2 V_T^2+\frac{\Omega^2}{\omega^2}
k_{\parallel}^2 V_T^2\right)^{-1}, \nonumber
\end{align}
where $V_T^2=3T_e/(m_e c^2)$ and $k^2=k_{\bot}^2+k_{||}^2$. 
The dashed curves in Fig. \ref{om2} (e) and (f) demonstrate that the frequency of thus defined upper-hybrid mode, getting into the Cherenkov resonance with the beam, rises as the plasma temperature increases resulting in the wrong position of the growth rate maximum line in the wavenumber space. Thus, even nonrelativistic temperatures of plasma electrons cannot be correctly described by fluid models in the regime of interest.

\subsection{The impact of superthermal electrons}

The impact of kinetic effects on unstable spectra becomes much stronger in non-Maxwellian plasmas with energetic tails of superthermal electrons. As Fig. \ref{r2} (a), (b) and (c) show, the rise of the momentum spread $\Delta p^{(e)}$ characterizing the temperature of core electrons not only narrows the spectrum of oblique instabilities, but significantly decreases the growth rate of the fastest longitudinal waves. In analogy with the Maxwellian plasma, let us analyze the real and imaginary parts of wave frequency (Fig. \ref{r2} (d) and (e)) along the $k$-lines of the growth rate maximum. Here, the effect of frequency decrease for the upper-hybrid mode in a hot plasma becomes more pronounced. For the momentum spread $\Delta p^{(e)}=0.3$, the magnetic contribution to the dispersion of this mode is almost not visible, the real frequency is localized near the plasma frequency (Fig. \ref{r2} (e)), and the line of the Cherenkov resonance transforms to the straight line $k_{||}=1/v_b$ (Fig. \ref{r2} (f)).

We recall that in the trapping regime even a small change in the growth rate of the most unstable mode should result in the reduction of the power pumping to the turbulence by the beam ( $P\propto \Gamma^5$). Our analysis shows in particular that if plasma heating is accompanied by the formation of strongly non-Maxwellian distribution of plasma electrons with the high enough core temperature $\sim 10$ keV ($\Delta p^{(e)}=0.2$) and the energetic superthermal tail, the  growth rate of the fastest two-stream instability is reduced by a factor of 1.5, which entails the reduction of the heating power by almost the order of magnitude 
($1.5^5\approx 7.6$).

\section{Summary}
We perform the first kinetic calculations of the growth rate for the instability of a hot electron beam in a hot magnetized plasma in the framework of the general linear theory based on the relativistic Vlasov equation. The stabilizing role of thermal effects in plasmas is investigated for the Maxwellian and for strongly nonequilibrium distributions containing intense tails of superthermal electrons typical for laboratory beam-plasma experiments. It is shown that nonrelativistic electron temperatures in the Maxwellian plasma can stabilize oblique instabilities only, while the tail formation in the nonmaxwellian plasma has an impact on the whole unstable spectrum and can substantially reduce the efficiency of turbulent plasma heating.

\begin{acknowledgments}
This work is supported by the Ministry of education and science of Russia (projects 14.B37.21.0750, 14.B37.21.1178, 14.B37.21.0784 and 8387), Russian Foundation of Basic Research (grants 12-02-31696,
11-02-00563, 11-01-00249) and the President
grant SP-1289.2012.1.
\end{acknowledgments}


\end{document}